# Results of a Collective Awareness Platforms Investigation

Giovanna Pacini[1] and Franco Bagnoli[1,2]

[1]Department of Physics and Astronomy and
CSDC-Center for the Study of Complex Dynamics, University of Florence,
Via G. Sansone 1, Sesto Fiorentino, Florence, Italy
[2]INFN, sez. Firenze
giovanna.pacini@unifi.it;franco.bagnoli@unifi.it

**Abstract.** In this paper we provide two introductory analyses of CAPs, based exclusively on the analysis of documents found on the Internet. The first analysis allowed us to investigate the world of CAPs, in particular for what concerned their status (dead or alive), the scope of those platforms and the typology of users. In order to develop a more accurate model of CAPs, and to understand more deeply the motivation of the users and the type of expected payoff, we analysed those CAPs from the above list that are still alive and we used two models developed for what concerned the virtual community and the collective intelligence.

## 1   CAPs: brief review

Collective Awareness Platforms are important crowdsourcing instruments that may promote cooperation, emergence of collective intelligence, participation and promotion of virtuous behaviours in the fields of social life, energy, sustainable environment, health, transportation, etc. [1,2].

CAPs do not obey in general to the usual market dynamics: they are developed by volunteers or after a public support (namely, EU projects). Also the participation of the public in CAPs is not due to an immediate return, and there are several motivations, exposed in the following, whose lack of analysis may lead to the failure of the CAP, with an evident waste of effort and public funding.

The core of our investigation is that of examining the motivations for the participation in CAPs based on a model of the individual user based on what is known of human behaviour beyond rationality: human heuristics, emotional components, peers and group influence. In particular, we shall analyse the role of payoff (which in general depends non-linearly on the number of participants), incentives, motivations (reputation, emotional components) and community structure.

## 2 The analysis

We analyzed 70 CAPs selected from those financed by EU in the last calls and others involved in European projects. To seek support for modeling the behavior of CAPs the survey followed the principle of group-specific purposive sampling. The CAPs under examination are extremely varied and therefore we tried to identify a limited set of dimensions to be investigated. The main points that we would like to study are:

- The subject of the action of the CAP: which field/problem/need this platform is addressing.
- The health state of the CAP: is it alive, dead, completed or failed?
- Number of participants, kind of community/group/hierarchical structure that the CAP is promoting.
- Messages and communications among participants, communication network.
- Role and structure of the expected payoff from the point of view of users.

### 2.1 Applicative field, status and target

We divided the CAPs according to their target fields, as shown in Fig. 1. Sustainability, ITC and sociology cover almost 60% of the total.

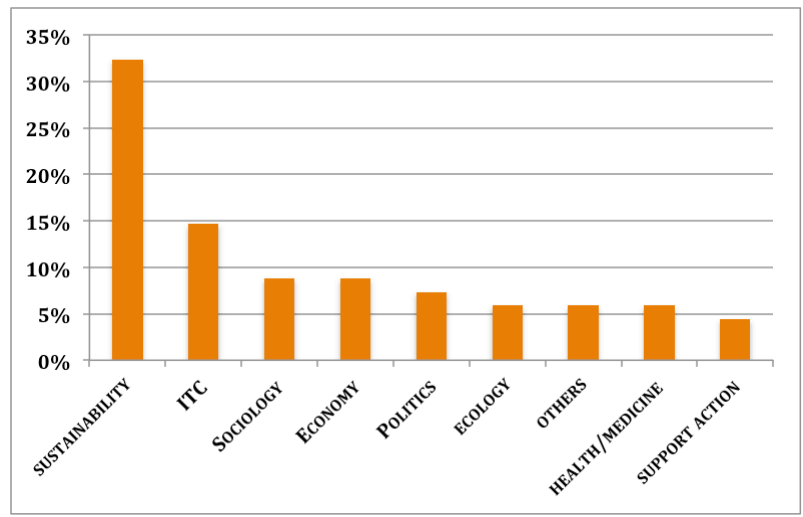

**Fig. 1.** Application field of CAPs

For what concerns the activity level of CAPS we found that the 75% of them are alive after at least three years from opening. Some of them may have moved their activity to other media (such as Facebook). What is remarkable (and will be the subject of a further investigation) is that inactive CAPs are, almost all, platforms developed within European projects.

An important aspect of this study is the evaluation of CAPs audience, intended both as a number and as a type of user. For what concerns the geographical target, we found that 50% have a worldwide audience, 20% a European one and the others are devoted to local targets. We divided the audience by category of users that may be involved in a CAP; almost half of the CAPS analyzed are addressed to citizens and about a 20% are dedicated to researchers.

### 2.2 Social media impact

It is very difficult to estimate the number of users from the data obtainable from web site. In many cases we collected data from Facebook and Twitter as reported collectively in Fig. 2.

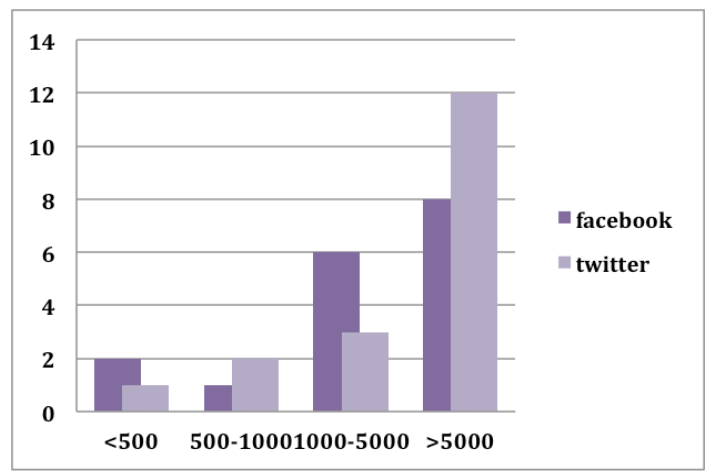

**Fig. 2.** Social media impact. In the horizontal axis there is the number of likes/followers; in the vertical axis the number of CAPS.

## 3 Second analysis

In order to develop a more accurate model of CAPs, and to understand more deeply the motivation of the users and the type of expected payoff, we further analysed those CAPs from the above list that are still alive. Our results are based exclusively on the analysis of documents found on the Internet.
We tried to understand why a user should use a CAP and we highlighted some reasons (there may be several reasons for each CAP). The results are reported in Fig. 3.

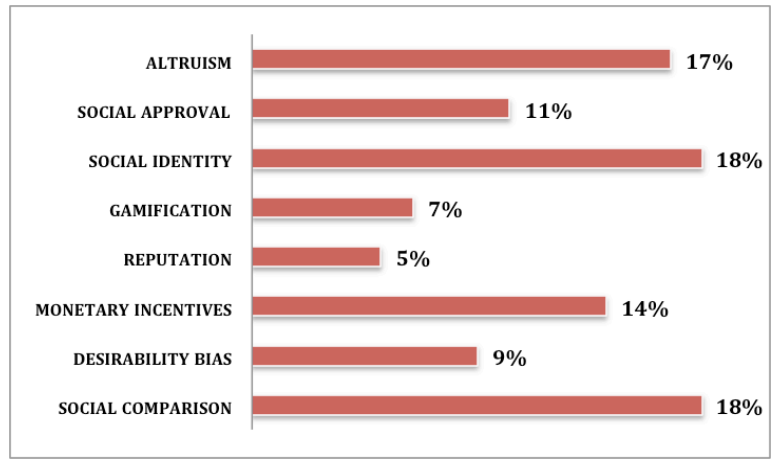

**Fig. 3.** Motivations of CAPs.

Let us now examine the type and scaling of payoff, i.e., the expected return for an user investing time and maybe money into a CAP. We can suppose four different scenarios, see Fig. 4.

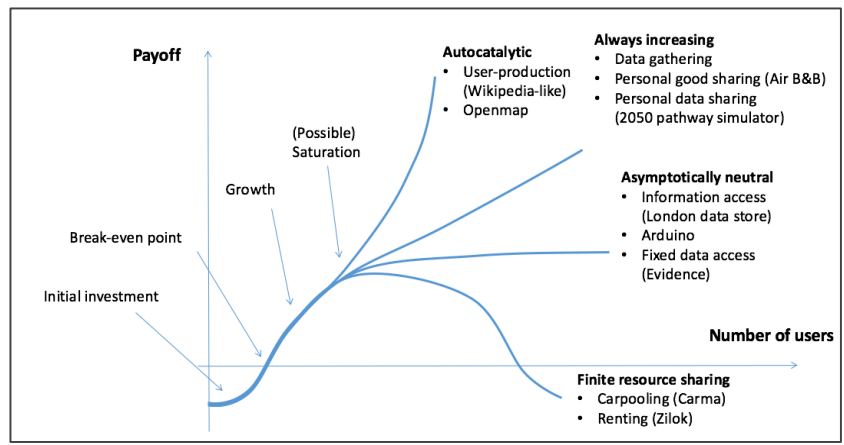

**Fig. 4.** Payoff versus number of users for different scenarios.

The first one is the autocatalytic one, which can be expected if the value of a CAP is given by the direct user-production (Wikipedia for instance). In this case, the more the users the more the payoff. The second scenario (always increasing) is similar, except that user participate in cataloguing and searching data, not in their production. It is the case for instance of AirB&B. The third scenario, asymptotically neutral, is given by CAPs that provide access to static pieces of information. After that user contributed, for instance by discussing and furnishing support, they may expect to receive a return which does not depend on the number of users. The final scenario, finite resource sharing, is typical of CAPs offering tools for accessing a finite resource, for instance alerting about free park slots. In this case there is an optimum in the number of users, after which the payoff decreases.

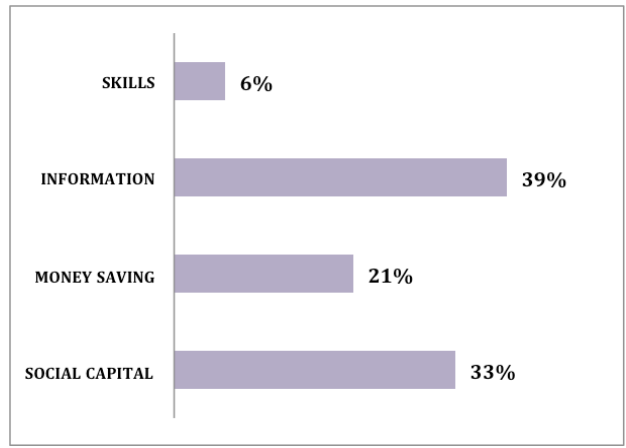

**Fig. 5.** Type of payoff

This payoff scaling does not mean that users could not be interested in accessing a given resource. As shown in Fig. 5, most of payoff (39%) is in form of information, money saving for 21 %, social capital for 33% and skill for 6%. As typical for this kind of resource, the payoff mainly increases with the number of users (54%), see Fig. 6.

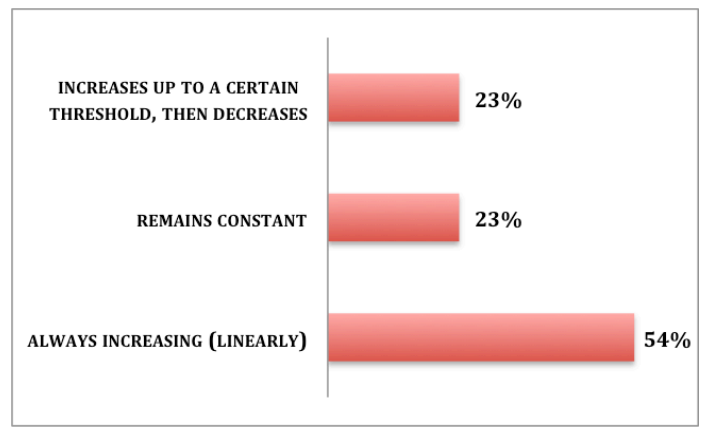

**Fig. 6.** Payoff scaling.

### 3.1 CAPs as virtual communities

Virtual communities are well know and well studied starting probably from 1985 by Rheingold [3]. We take the definition of *virtual community* from Porter [4]. A virtual community can be defined as

- An aggregation of individuals or business partners
- Who interact around a shared interest,
- Where the interaction is at least partially supported and/or mediated by technology and guided by some protocols or norms.

So we can look at a CAP as a type of virtual community and try to analyse them within the framework of the models by Markus [5] and Malone [6].

**Characterization by Markus.**

Markus [5] presented a classification of the virtual communities based on the community's social, professional or commercial orientation. She explained:

> *This characterization is based on the existing divisions but also attempts to provide a framework for establishing divisions that are as clear cut as possible, without potential for overlaps.*

The CAPs examined, as shown in Fig. 7 only for the first level, have mostly a social orientation, which is understandable with the general purpose of collective awareness platforms.

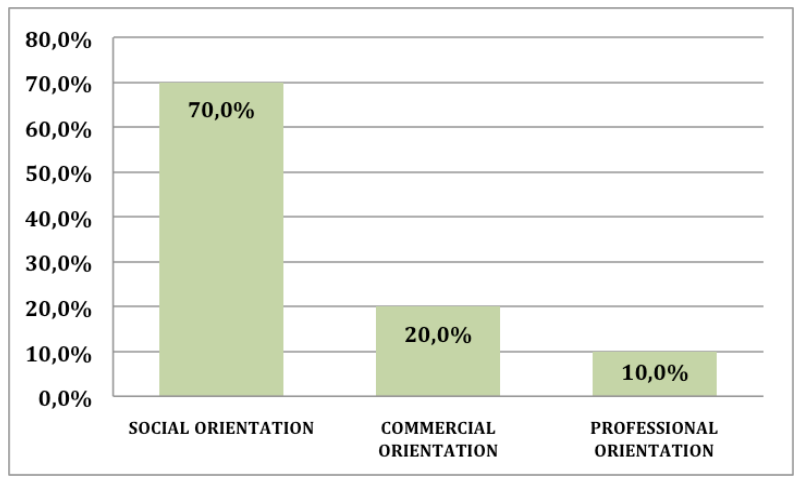

**Fig. 7.** Orientation of CAPs as virtual communities.

**Characterization by Malone.**

Another very interesting work by Malone *et al*. [6] is based on the concept collective intelligence. Their idea was to describe every system (in particular an IT System) in terms of blocks of collective intelligence, so that one can speak of the "genome" of a collective intelligence system. The identification of the genome is based on four questions:

- What is being done?
- Who is doing it?
- Why are they doing it?
- How is it being done?

The first level can be further specified.

**What?** This is the first question to be answered for any activity. It is the mission or goal or simply the task. The task can be to *Create* (make something new) or *Decide* (evaluate and select alternative).

**Who?** The question is about who undertakes an activity. Possible answers are: *Hierarchy*, (someone in authority assigns a particular person or group to perform a task) or *Crowd*, (anyone in a large group who chooses to do so).

**Why?** This question deals with incentives, the reason for which people take part in the activity. What motivates them? What incentives are at work? The possible answers are: *Money*, where participants earn money from the activity, *Love*, in the sense of intrinsic enjoyment of the activity, the opportunity to socialize, the idea of contributing to something larger than themselves, *Glory* (or reputation), which is the recognition of themselves among their community.

In this our first analyze we did not investigate the question "how" and tried to apply the classification to different roles inside a CAP (user, owners, researcher and so on). The results are shown in Fig. 8.

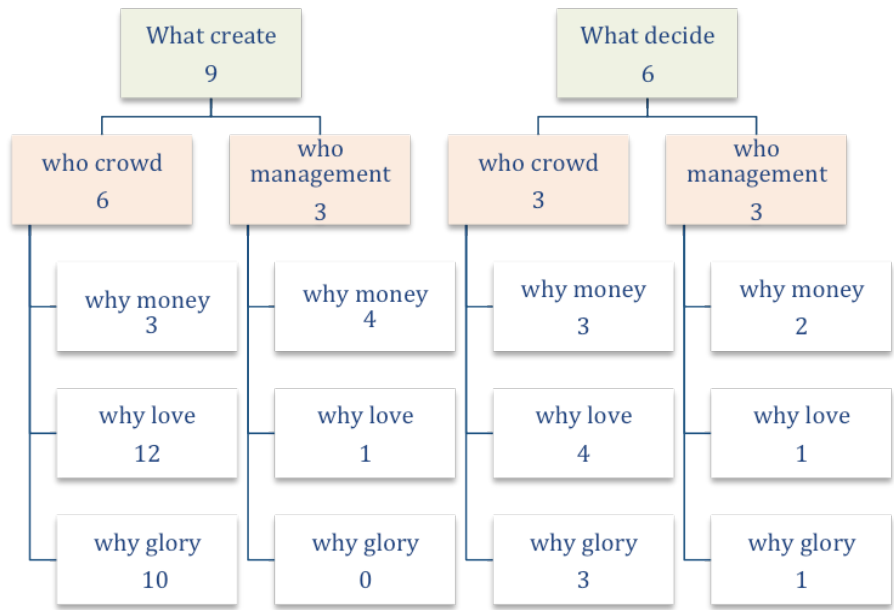

**Fig. 8.** Genes of collective intelligence of CAPs. The numbers represent how many times we have found the genes.

## 4     Conclusions

We performed two studies on the aspects that may influence the performances and health status of CAPs. The main lessons are:

- Almost two-third of CAPs are developing applications on sustainability, ITC and sociology.
- Most CAPs do not report on their stakeholders outreach and audiences and therefore it is impossible to evaluate their impact.
- As for potential user involvement, almost half of the CAPs analysed are addressed to citizens and one fifth are dedicated to researchers.
- Most of alive CAPs share information as the payoff for users, and since this resource is in general furnished by users, the expected payoff per user is constant or increases with the number of users themselves, and this is an indicator of a possible further increase of the CAPs audience.

**Acknowledgements.** The support of the European Commission under the FP7 Programme Collective-Awareness Platforms under the Grant Agreement no ICT-611299 for the SciCafe2.0 is gratefully acknowledged.